\documentclass[aps,prb,superscriptaddress,twocolumn,showpacs,amsmath,amssymb,floats]{revtex4-1}
\usepackage{graphicx}% Include figure files
\begin{document}
\title{Angle-resolved photoemission spectroscopy study of HgBa$_{2}$CuO$_{4+\delta}$}
\author{I. M. Vishik}
\affiliation {Massachusetts Institute of Technology, Department of Physics, Cambridge, MA, 02139, USA}
\author{N. Bari\v{s}i\'{c}}
%\affiliation{Service de Physique de l’Etat Condens\'{e}, CEA-DSM-IRAMIS, F 91198 Gif-sur-Yvette, France}
\affiliation{University of Minnesota, School of Physics and Astronomy, Minneapolis, MN 55455, USA}
\affiliation{Institute of Solid State Physics, Vienna University of Technology, Vienna, 1040, Austria}
\author{M.K. Chan}
\affiliation{University of Minnesota, School of Physics and Astronomy, Minneapolis, MN 55455, USA}
\author{Y. Li}
\affiliation{Peking University, International Center for Quantum Materials, School of Physics, Beijing,  100871, China}
\affiliation{Collaborative Innovation Center of Quantum Matter, Beijing, 100871, China}
\author{D. D. Xia}
\altaffiliation[Current address:]{School of Pharmaceutical Engineering, Shenyang Pharmaceutical University,Shenyang, 110016, China}
\affiliation{University of Minnesota, School of Physics and Astronomy, Minneapolis, MN 55455, USA}
\affiliation{State Key Laboratory of Inorganic Synthesis and Preparative Chemistry, College of Chemistry, Jilin University, Changchun, 130012, China}
\author{G. Yu}
\affiliation{University of Minnesota, School of Physics and Astronomy, Minneapolis, MN 55455, USA}
\author{X. Zhao}
\affiliation{University of Minnesota, School of Physics and Astronomy, Minneapolis, MN 55455, USA}
\affiliation{State Key Laboratory of Inorganic Synthesis and Preparative Chemistry, College of Chemistry, Jilin University, Changchun, 130012, China}
\author{W. S. Lee}
\affiliation {Stanford Institute for Materials and Energy Sciences, SLAC National Accelerator Laboratory, 2575 Sand Hill Road, Menlo Park, CA 94025, USA}
\affiliation{Geballe Laboratory for Advanced Materials, Departments of Physics and Applied Physics, Stanford University, Stanford, CA 94305, USA}
\author {W. Meevasana}
\affiliation{Suranaree University of Technology, School of Physics, Muang, Nakhon Ratchasima, 30000, Thailand}
\author{T. P. Devereaux}
\affiliation {Stanford Institute for Materials and Energy Sciences, SLAC National Accelerator Laboratory, 2575 Sand Hill Road, Menlo Park, CA 94025, USA}
\affiliation{Geballe Laboratory for Advanced Materials, Departments of Physics and Applied Physics, Stanford University, Stanford, CA 94305, USA}
\author{M. Greven}
\affiliation{University of Minnesota, School of Physics and Astronomy, Minneapolis, MN 55455, USA}
\author{Z. X. Shen}
\affiliation {Stanford Institute for Materials and Energy Sciences, SLAC National Accelerator Laboratory, 2575 Sand Hill Road, Menlo Park, CA 94025, USA}
\affiliation{Geballe Laboratory for Advanced Materials, Departments of Physics and Applied Physics, Stanford University, Stanford, CA 94305, USA}

\date{\today}% It is always \today, today,

\begin{abstract}
HgBa$_{2}$CuO$_{4+\delta}$ (Hg1201) has been shown to be a model cuprate for scattering, optical, and transport experiments, but angle-resolved photoemission spectroscopy (ARPES) data are still lacking owing to the absence of a charge-neutral cleavage plane.  We report on progress in achieving the experimental conditions for which quasiparticles can be observed in the near-nodal region of the Fermi surface.  The \textit{d}-wave superconducting gap is measured and found to have a maximum of 39 meV.  At low temperature, a kink is detected in the nodal dispersion at approximately 51 meV below the Fermi level, an energy that is different from other cuprates with comparable \textit{T}$_c$.  The superconducting gap, Fermi surface, and nodal band renormalization measured here provide a crucial momentum-space complement to other experimental probes.
\end{abstract}

\maketitle

Hg1201 is a cuprate whose structural simplicity and low residual resistivity makes it an ideal compound for many experiments including charge transport\cite{Barisic:UniversalSheetResistance}, Raman \cite{LeTacon:TwoEnergyScales_Raman_Hg1201,Li:RamanHg1201_2012,*Li:DopingDepPhotonScatteringResponance}, NMR \cite{bobroff:NMR_Hg1201,*rybicki:spatialInhomogeneitiesNMR2009,*Haase:2ComponentSpinSusceptibilityHg1201}, thermodynamics \cite{Gribic:MicrowaveHg1201_2009,Xia:anisotropyParameter_2012,Hardy:PressureHg1201_2010,*wang:strainHg1201}, and neutron\cite{Li:UnusualMagneticOrderHg1201,*Li:MagneticModeHg1201,*Yu:MagneticResonanceModeHg1201_2010,*Li:MagneticOrderPGHg1201_2011,*Li:MagneticVortexLatticeHg1201_2011,*Li:TwoIsingLikeMagneticExcitations} and x-ray scattering \cite{dAstuto:PhononHg1201, Lu:ChargeTransferExcitationsModelSC}.  It has a simple tetragonal single-layer crystal structure (P4$/$mmm), and oxygen dopants reside in the Hg layer, relatively far from the CuO$_2$ planes, minimizing disorder effects \cite{Eisaki:ChemicalInhomogeneity,Barisic:DemonstratingModelNatureHg1201}. In fact, it is a model cuprate for gaining quantitative information from transport and optical measurements \cite{Barisic:DemonstratingModelNatureHg1201,Barisic:UniversalSheetResistance,vanHeumen:opticsThermodynamicsHg1201,Mirzaei:SpectroscopicFL_PG_Hg1201,Doiron:HallSeebeckNernst}.  Additionally, quantum oscillations have recently been reported in underdoped Hg1201, attesting to the long mean free path and confirming the universality of small Fermi pockets in the field-induced resistive state \cite{barisic:UniversalQO}.  The highest superconducting \textit{T}$_c$ was reported in a related triple-layer compound \cite{Gao:Highest_Tc}, so in addition to offering general insight on the cuprates, the study of Hg1201 may provide a perspective on how to maximize \textit{T}$_c$\cite{Raghu:ReservoirLayer}.

However, there have not yet been peer reviewed ARPES studies of Hg1201 in the literature.  The reason for this is that the cleaved surfaces are not ideal, which makes it difficult to obtain useful spectra.  Here, we present the progress we have made in obtaining the optimal experimental conditions to study Hg1201 with ARPES, which allowed us to quantitatively measure the near-nodal electronic structure of this material.   We find that data quality strongly hinges on selecting the proper experimental conditions.  With the experiment optimized to achieve the sharpest near-nodal spectra, we are able to estimate  the Luttinger volume, measure the \textit{d}-wave superconducting gap near the node, and observe a nodal renormalization feature near 50 meV.
%Information about the Fermi surface shape and volume, the superconducting gap, and band renormalization gained from ARPES is important for the interpretation of other experiments for which Hg1201 is a model compound.

\section{Experimental conditions}
\begin{figure}[!]
\includegraphics [type=jpg,ext=.jpg,read=.jpg,clip, width=2.75 in]{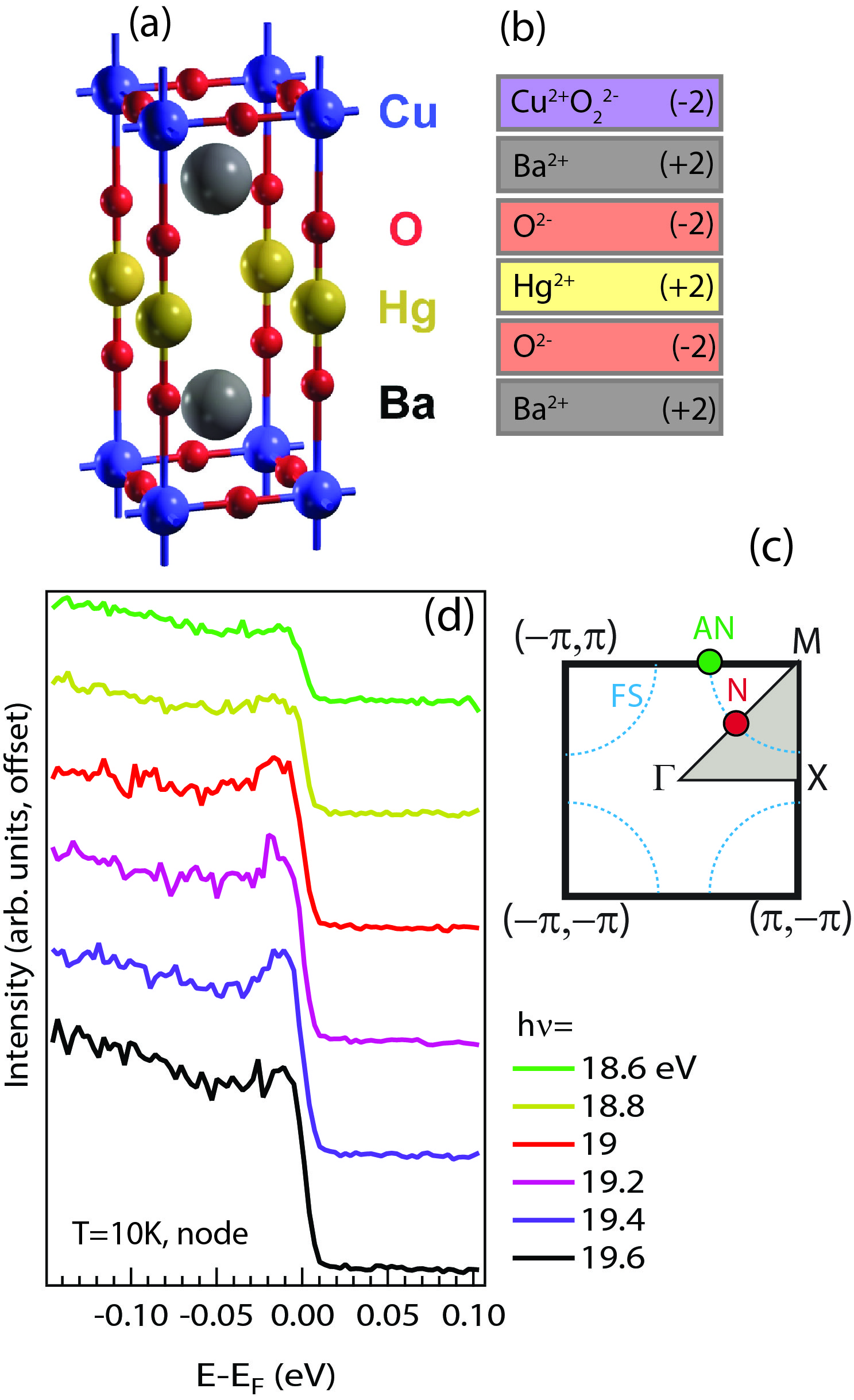}
\centering
\caption[Crystal structure of Hg1201 and hn dep of nodal EDC]{\label{Fig 1: Crystal structure and hn dep} (a) Schematic crystal structure of Hg1201 for simplicity drawn without oxygen interstitials in the Hg layers (top).  Image from Ref. \onlinecite{Barisic:UniversalSheetResistance}.  (b) charges of each layer in the crystal structure, excluding oxygen interstitials. (c) 2D projection of tetragonal Brillouin zone (BZ) with high symmetry points labeled.  Schematic of Fermi surface (FS) is shown by blue dashed lines. Red dot marks the nodal (N) momentum, where the Fermi surface intersects the BZ diagonal ((0,0)-($\pi$,$\pi$) line), and green dot marks the antinodal (AN) momentum, where the FS intersects the BZ boundary.  (d) Photon energy dependence of EDCs at nodal momentum (red dot in (c).  Data taken on a single sample at SSRL with cuts along BZ diagonal.}
\end{figure}

The schematic crystal structure of Hg1201 is shown in Fig. \ref{Fig 1: Crystal structure and hn dep}(a).  It can be seen that the  structure lacks a neutral cleavage plane, as there is nowhere to slice the unit cell in a fashion that would yield identical atomic planes on both sides of the cleave.  Additionally, the Hg1201 crystal structure, adjacent layers have opposite charges (Fig. \ref{Fig 1: Crystal structure and hn dep}(b)), so any pure termination would produce a polar catastrophe \cite{Nakagawa:PolarCatastrophe} (electric potential increasing to infinity) which must be mitigated either by self-doping or by a mixed-termination.  Scanning surface microscopies of Hg1201 have indicated that the topography of cleaved surfaces consists of 4nm step edges with flat terraces extending hundreds of nanometers \cite{Hubler:Hg1201_scanning_spectroscopies}.  Core level x-ray photoelectron spectroscopy (XPS) suggests that the surface termination is near the Hg layer because the Hg:Cu and Hg:Ba intensity ratios normalized by the XPS sensitivity to those elements was larger than expected from the chemical formula\cite{YuanLi:Thesis}.

Single crystals were grown by the two-step flux method by which a maximum \textit{T}$_c$ onset of 96.5K can be achieved \cite{Zhao:CrystalGrowthHg1201}.  In preparing Hg1201 samples for ARPES experiments, care was taken to ensure electrical conductance between the sample and the sample post.  Samples were glued onto the copper sample post using EPO-TEK H21D silver epoxy (Epoxy technology Inc).  While this silver epoxy provides adequate conduction between the copper post and most compounds, Hg1201 samples were not found to be properly grounded to the post after this step.  This may be due to the epoxy reacting with the Hg1201 surface.  Thus, the silver epoxy provides only mechanical adhesion in our experiments.  For conduction, silver paint (Dupont 4299N) was applied to the side of the sample and the copper post and cured at room temperature.  Conduction between the top of the sample and the copper post was confirmed.  Care was also taken to maximize the probability of a good cleaved surface.  A pre-cut was made on the side  of the sample using a surgical razor blade, parallel to the a-b face.  This was to ensure cleaving at a designated location, rather than at inclusions and imperfections as would be the tendency without a pre-cut \cite{Barisic:DemonstratingModelNatureHg1201}.   This is one of the essential steps that enabled successful ARPES measurements.

We studied nearly-optimally-doped Hg1201 (\textit{T}$_c$=95K, determined at the transition midpoint).   Data were taken at SSRL at 10K with a Scienta R4000 analyzer and 10 meV energy resolution.  Samples were cleaved in-situ at a pressure better than 5$\times$10$^{-11}$ Torr.

Experiments were attempted with a 7eV laser, near 55 eV at beamline 10.0.1 of the Advanced Light Source (ALS), and near 19 eV at beamline 5-4 at Stanford Synchrotron Radiation Lightsource (SSRL).  The latter experimental condition was found to yield the best spectra, and the quality of the measured spectra depended sensitively on the experimental conditions.  Fig. \ref{Fig 1: Crystal structure and hn dep}(d) shows energy distribution curves (EDCs) at the Fermi crossing momentum, k$_F$, at the node taken with several photon energies between 18.6 and 19.6 eV.  The amplitude of the quasiparticle peak relative to the background varies rapidly with the choice of photon energy, even within this narrow range.  Quasiparticle peaks were clearly observed for 19-19.4 eV photon energy and very weak for 18.6, 18.8, and 19.6 eV photon energies.  The nodal quasiparticle peak was found to be most pronounced for 19.4 eV photon energy and $\Gamma$-M cut geometry.  For this cut geometry, the polarization of the beam at SSRL is 45$^\circ$ from the Cu-O bond direction.  Note that because Hg-1201 is tetragonal, the M point is the Brillouin zone corner, and $\Gamma$-M cuts are along the (0,0)-($\pi$,$\pi$) line.

%WRITE SOMETHING ABOUT MATRIX ELEMENTS FOR NON-EXPERTS

\section{Momentum dependence}

\begin{figure*}[!]
\includegraphics [type=jpg,ext=.jpg,read=.jpg,clip, width=4.75 in]{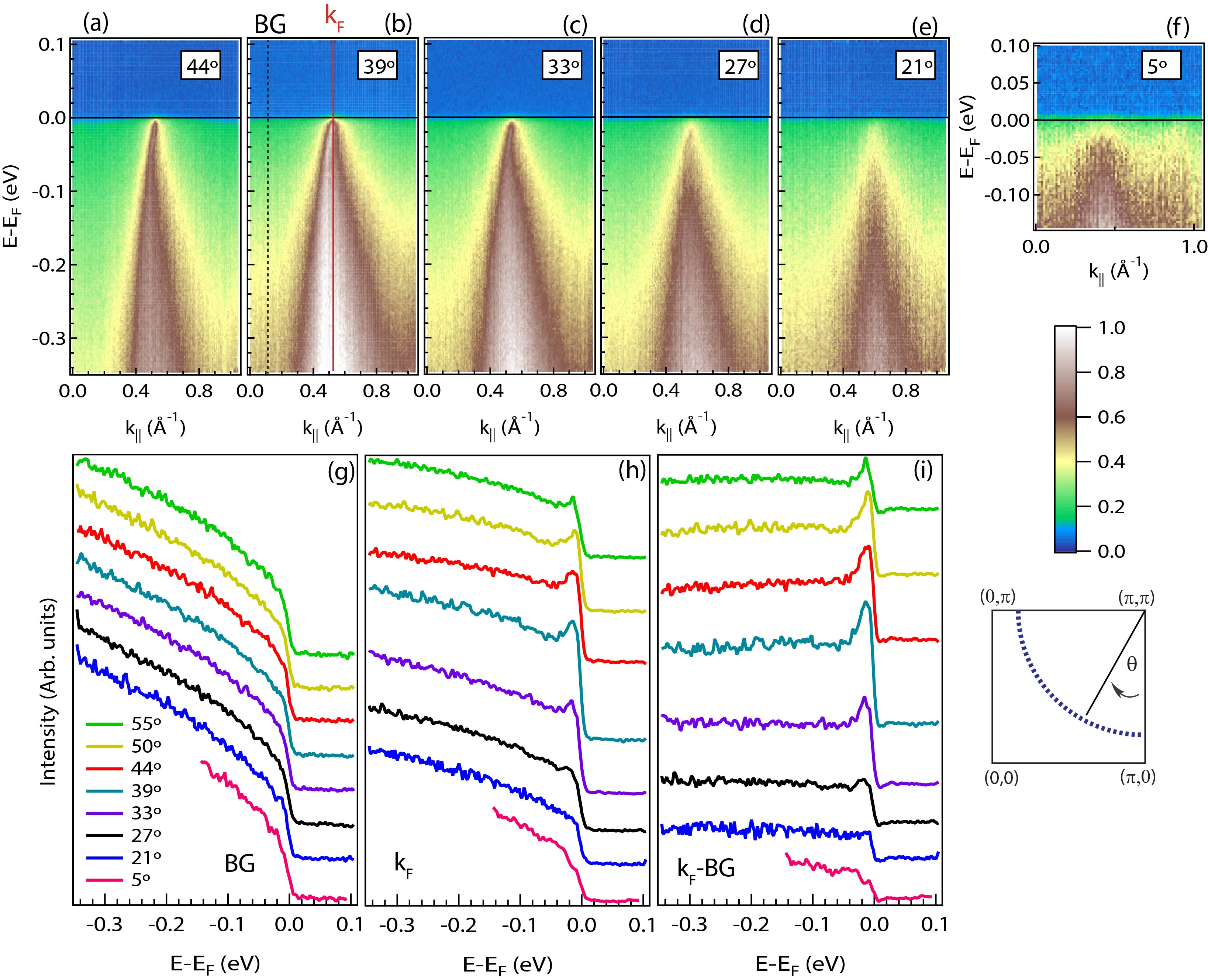}
\centering
\caption[Near nodal cuts]{\label{Fig 3: Near-nodal cuts Hg1201} (a)-(f) Image plots of near-nodal cuts.  All images have the same linear color scale.  Color scale and Fermi surface angle $\theta$ are defined below (f).  Color scale is in arbitrary units.  Data taken with 19eV photons, cuts parallel to $\Gamma$M, and T$=$10K.  Red vertical line in (b) marks Fermi momentum (k$_F$) and black dashed line marks momentum of background EDC. (g) Background EDCs, taken at momentum indicated in (b). (h) EDC at k$_F$. (i) EDC at k$_F$ with background EDC subtracted.}
\end{figure*}

Using the optimal experimental configuration, electron states near the node are accessible, but the spectral intensity is very small at the antinode, likely because of extrinsic reasons.  Fig. \ref{Fig 3: Near-nodal cuts Hg1201} shows momentum dependence of spectra.  The band is most pronounced near the node in this experimental configuration, and spectral intensity diminishes away from the node.  By $\theta=$21$^\circ$, quasiparticles are no longer observed. EDCs at k$_F$ are shown in Fig. \ref{Fig 3: Near-nodal cuts Hg1201}(h).  All spectra have a strong energy-dependent background, shown in Fig. \ref{Fig 3: Near-nodal cuts Hg1201}(f), which is identical for all cuts.  Previous studies have suggested that this ARPES background is attributed to photoelectrons which scattered inside the sample and lost their momentum information prior to being photoemitted \cite{Kaminski:ARPES_Background}.  In Hg1201, additional contribution to the background may come from photoelectrons scattering from surface step-edges.  Subtracting this background EDC highlights the quasiparticle peaks, as shown in Fig. \ref{Fig 3: Near-nodal cuts Hg1201}(i).  Similar methods have been used to discern spectral features buried beneath a large background in other cuprates \cite{Hashimoto:BGsubtBi2201}.

The apparent decrease of cross-section away from the node is likely not intrinsic, as most cuprates with comparable \textit{T}$_c$ show antinodal quasiparticles at optimal doping with proper experimental conditions \cite{Lu:YBCO_ARPES,Vishik:QPI_ARPES,Lee:Tl_cuprate}. Matrix element effects are generally one reason spectral intensity may be unobservable.  In ARPES experiments, the measured spectral intensity is modulated by a dipole matrix element term $|M|^2=|\langle \Psi_f|\textbf{A}\cdot \textbf{p}|\Psi_i \rangle|^2$ where \textbf{A} is the photon vector potential, \textbf{p} is the electron momentum, and $\Psi_f (\Psi_i)$ are wavefunctions of the final (initial) electron states.  Changes in the photon energy or polarization can affect the intensity of a band measured by ARPES and also which orbitals are highlighted \cite{Bansil:MatrixElements2004,Lindroos:MatrixElementsPolarizationFinalStates}.  For cuts along high-symmetry directions in the Brilliouin zone, certain light polarization can suppress a band in the ARPES spectrum based on its orbital character.  Final-state effects can completely suppress a band in the ARPES spectrum if there are no final states to excite into with the chosen photon energy.  Further exploration is needed to find an experimental configuration where the antinode has adequate cross section for Hg1201.   We note that quasiparticles near the node are pronounced only within a limited range of photon energies (Fig. \ref{Fig 1: Crystal structure and hn dep}(d)), which is one of the reasons that ARPES data on Hg1201 were not available previously.  Optimizing the experiment for the antinode will likely require similar careful exploration over parameter space--photon energy, polarization, cut geometry.

\begin{figure}[!]
\includegraphics [type=jpg,ext=.jpg,read=.jpg,clip, width=3 in]{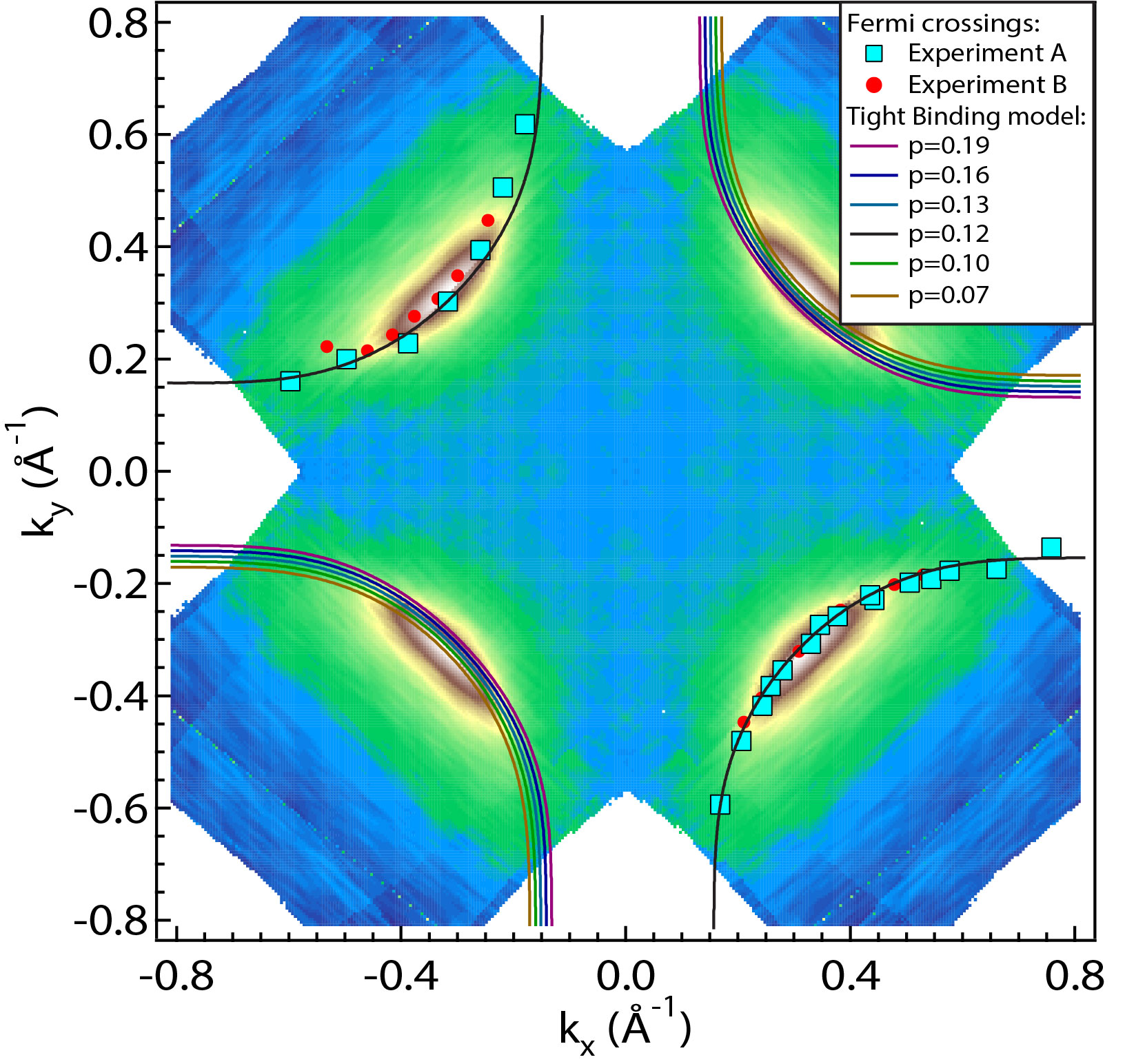}
\centering
\caption[Fermi surface mapping Hg1201]{\label{Fig 4: Fermi surface mapping Hg1201} Fermi surface map, k$_F$ for each cut, and tight-binding model Fermi surfaces. Map is produced by integrating cuts within a 20 meV window centered at E$_F$. Four-fold symmetrization is applied to the color map, but not k$_F$ data. Red and blue symbols are Fermi crossings for two different experiments on two different batches of crystals.  For each cut, k$_F$ was determined from the peak position of the momentum distribution curve at E$_F$.  In the superconducting state, resolution effects produce finite intensity at E$_F$ near the node from which this determination can be made.  Experiment A was performed with 19eV photons and experiment B was performed with 19.4 eV photons.  Solid lines are tight-binding Fermi surfaces enclosing different volumes, with hopping parameters from Ref. \onlinecite{Das:Q0_collectiveModesHg1201_thry}.  For quadrants with k$_F$ data, Fermi surfaces are only shown for p$=$0.12.}
\end{figure}

Fig. \ref{Fig 4: Fermi surface mapping Hg1201} shows a Fermi surface color map and k$_F$ values for each cut.  Data taken with 19 eV (experiment A) and 19.4 eV (Experiment B) photons yielded a similar Fermi surface.  Data were taken on two opposite sides of the $\Gamma$ point on both sides of the node.  Four-fold rotational symmetry was assumed to fill the quadrants in which data were not taken in the color plot, but symmetrization was not applied to k$_F$ data (blue squares, red circles).  Fig. \ref{Fig 4: Fermi surface mapping Hg1201} also shows tight binding Fermi surfaces for hole dopings 0.07$<$p$<$0.19 with hopping parameters which provide the best fit to dispersions from first-principles calculations \cite{Das:Q0_collectiveModesHg1201_thry}:$(t,t',t'', t''')=(0.46, -0.105, 0.08, -0.02)$.

The chemical potential was adjusted to compare to the Fermi-crossings data (red and blue symbols in Fig. \ref{Fig 4: Fermi surface mapping Hg1201}).  For experiment A, the Fermi surface encloses an area of 1.47$\AA$$^{-2}$, which amounts to a hole doping of 11.9$\pm$1.2\%, using 1$+$p$=$2A$_{FS}$/A$_{BZ}$, where A$_{FS}$ (A$_{BZ}$) is the area of the Fermi surface (Brillouin zone). Experiment B yielded a hole doping of 11.6$\pm$0.7$\%$.  A lattice constant of 3.876 $\AA$ was used \cite{Wagner:Hg1201CrystalStructure_1993}.  We note that the correspondence between nominal doping and FS area is much better than in as-cleaved YBa$_2$Cu$_3$O$_{7-\delta}$ (YBCO), another cuprate which lacks a neutral cleavage plane.  In YBCO, a polar catastrophe is avoided by self-doping of the surface layer such that as-cleaved surfaces are very overdoped (p$\approx$0.3) \cite{Hossain:InSituDopingControlARPESYBCO}, and in-situ surface dosing with potassium is required to produce optimally-doped and underdoped YBCO surfaces for ARPES study.  The measured FS area in Hg1201 indicates that self-doping of the surface layer is less of an issue, which makes Hg1201 is a promising cuprate for comparing ARPES data to established bulk probes without the need for complex surface preparation.

Several caveats should be considered when interpreting the Fermi surface volume in Fig. \ref{Fig 4: Fermi surface mapping Hg1201}.  First, there is uncertainty due to the lack of observed Fermi crossings in the antinodal region.  The exact shape of the Fermi surface is not known because the antinodal segments have not been accessed experimentally.  Second, the lack of a neutral cleavage plane may cause the hole concentration on the surface to be different from the bulk and perhaps to vary slightly between cleaves.  It should also be noted that the agreement between Fermi surface area and nominal doping in the literature is mixed.  In La$_{2-x}$Sr$_x$CuO$_4$ (LSCO), the Fermi surface area follows Luttinger's theorem, while in Ca$_{2-x}$Na$_x$CuO$_2$Cl$_2$ (Na-CCOC) and Bi$_2$Sr$_2$CuO$_{6+\delta}$ (Bi-2201), the FS area increases more rapidly with nominal doping \cite{Hashimoto:DopingEvolutionBi2201,Yoshida:FS_evolution_LSCO}.  This has been associated with differences in how doping occurs in different cuprates.  In LSCO, the chemical potential appears to be pinned inside the charge transfer gap, while in Na-CCOC and Bi-2201, the chemical potential shifts with doping. Finally, we note that thermoelectric power measurements of powder samples found a maximum \textit{T}$_c$ of 98K in Hg-1201 corresponding to 15.7$\%$ hole doping and a \textit{T}$_c$ of 95K corresponding to 12.7$\%$ hole doping \cite{Yamamoto:TEP_Hg1201}, which is more consistent with our ARPES data.   Future doping-dependent ARPES studies of the FS in Hg1201 can be expected to clarify the interpretation of the FS area observed in this experiment.
%NEED T=10K LATTICE PARAMETER

\begin{figure}[h!]
\includegraphics [type=jpg,ext=.jpg,read=.jpg,clip, width=3.5 in]{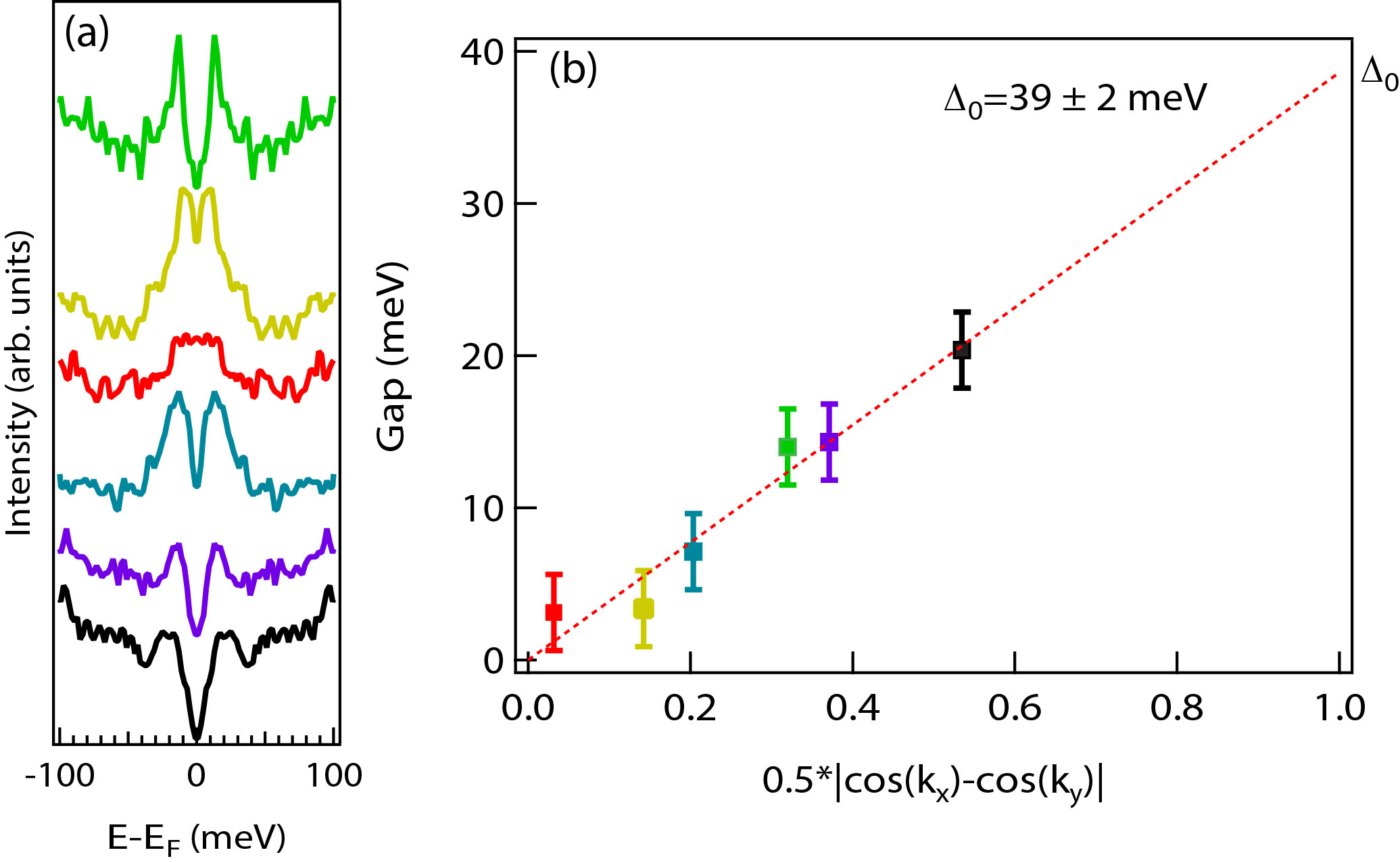}
\centering
\caption[Momentum dependence of gap, 10K]{\label{Fig 5: Momentum dependence of gap} Gap measurements, T$=$10K. (a) Symmetrized EDCs at k$_F$, using cuts in Fig. \ref{Fig 3: Near-nodal cuts Hg1201}.  Background EDCs (Fig. \ref{Fig 3: Near-nodal cuts Hg1201}(g)) have been subtracted.  EDCs are offset vertically. (b) Fitted gap at each momentum, plotted as a function of the simple \textit{d}-wave form.  Colors of EDCs in (a) correspond to colors of data points in (b) and to FS angles indicated in Fig. \ref{Fig 3: Near-nodal cuts Hg1201}(g)-(i).  Dotted line is linear fit, fixing y-intercept to zero. }
\end{figure}

With a favorable experimental configuration, we were able to measure the momentum dependence of the superconducting gap in Hg1201 using ARPES, as shown in Fig. \ref{Fig 5: Momentum dependence of gap}. Background EDCs were subtracted in order to emphasize the quasiparticle contribution for accurate extraction of the gap.  E$_F$ was determined from a polycrystalline gold sample which was electrically connected to the Hg1201 sample.  Spectra were symmetrized and EDCs at k$_f$ were fit to a minimal model convolved with the energy resolution of the experiment \cite{Symmetrization_Norman_model}.  Fitted gaps are plotted as a function of the simple \textit{d}-wave form, 0.5$|$$\cos$(k$_x$)$-$$\cos$(k$_y$)$|$.  Extrapolating to the antinode, assuming the gap function obeys a simple \textit{d}-wave form, gives an antinodal gap of 39$\pm$2 meV.  This is similar to the antinodal gap observed in Bi-2212 by ARPES at optimal doping \cite{Lee:twoGapARPES_TDep, Vishik:PNAS}, and also consistent with the doping-independent near-nodal gap slope observed in Bi-2212 0.076$<$p$<$0.19 \cite{Vishik:PNAS}.  This value of the superconducting gap ($\Delta$) is furthermore consistent with the scaling between $\Delta$ and the magnetic resonance mode energy explored in Ref. \onlinecite{Yu:UniversalRelationshipResonanceMode}. For Hg1201, Raman spectroscopy shows a peak in the B$_{1G}$ channel corresponding to a 41 meV antinodal gap \cite{Li:DopingDepPhotonScatteringResponance}, and tunneling experiments have reported a \textit{d}-wave gap with a maximum of 33 meV\cite{Wei:TunnelingHg1201}.

\section{Nodal dispersion analysis}

\begin{figure}[!]
\includegraphics [type=jpg,ext=.jpg,read=.jpg,clip, width=3.25 in]{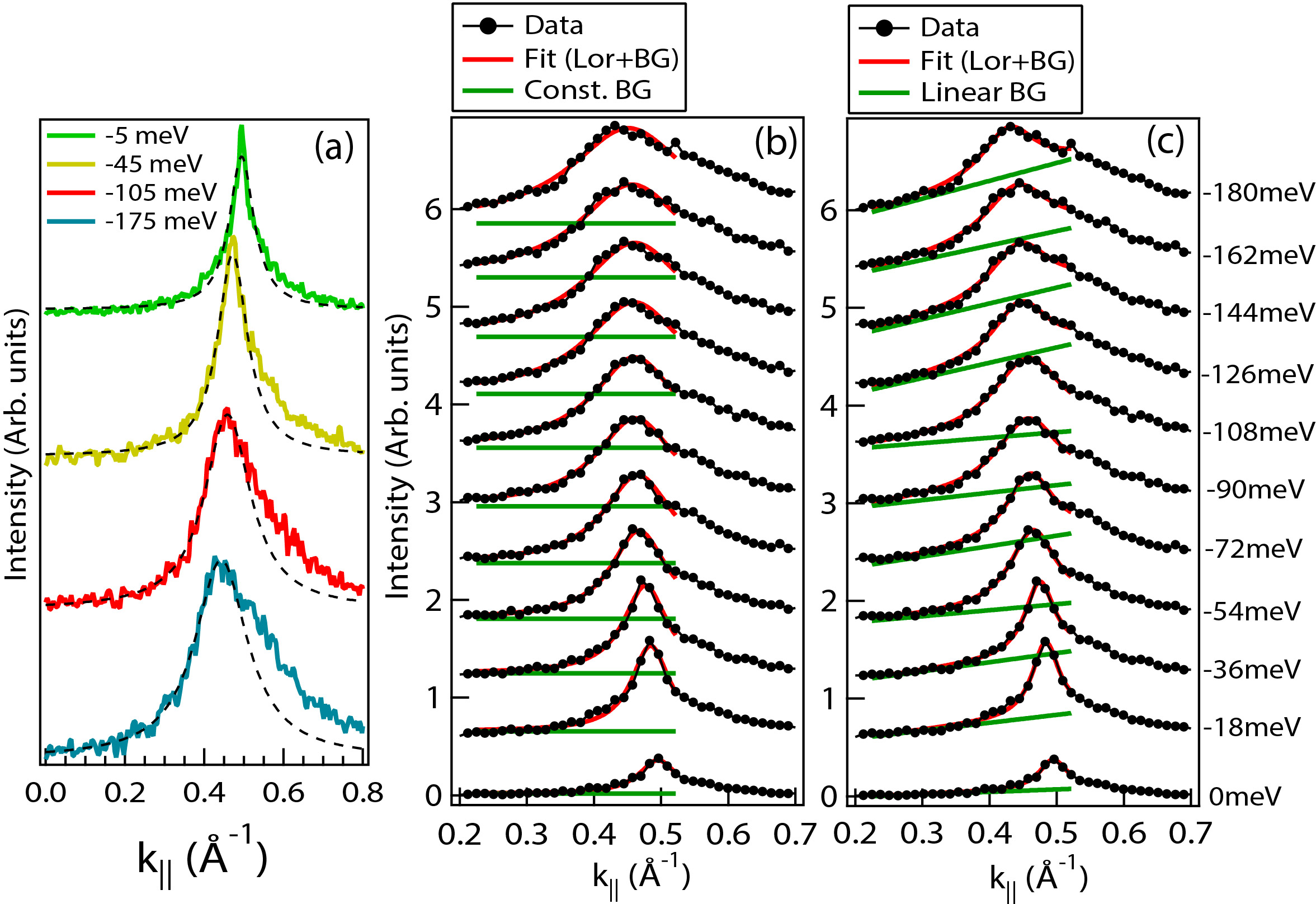}
\centering
\caption[MDCs at node]{\label{Fig 6: MDCs at node} MDCs at node. Data taken with 19.4eV photons.(a) Selected MDCs at indicated energies.  Black dashed lines are Lorentzian fits to the left-hand side of each peak. (b)-(c) Fitting of MDCs separated by 18 meV in energy to a Lorentzian plus background (red curves).  Green curves correspond to background, which is taken to be constant for (b) and linear (a$+$b\textbf{k}) for (c).  The left and right limits of the red and green curves indicate the momentum range of the fit.  Each MDC (black) is offset from the previous by 0.6 (arbitrary units). }
\end{figure}

Fig. \ref{Fig 6: MDCs at node} shows momentum distribution curves (MDCs) taken at the node.  Selected MDCs are shown in panel (a), and they have certain peculiarities which were observed in every experiment on Hg1201.  First, the MDCs near E$_F$ deviate from a Lorentzian lineshape, with extra weight in the tails, such that the peak height and width cannot be simultaneously captured.  Moving to higher binding energy, the MDCs become increasingly asymmetric, with extra weight on the side of the peak further from the $\Gamma$ point. Every sample studied showed the same asymmetry.  The MDC asymmetry might reflect interesting physics, such as a momentum-dependent self energy due to correlation effects\cite{Brouet:CoboltateARPESAsymmetricLineshape}.  However, given the lack of a neutral cleavage plane, surface electric fields might be responsible \cite{Hansen:SurfaceFields}.  Because of this MDC lineshape, using the usual fitting procedure of a Lorentzian peak plus a constant background does not yield the correct peak position at higher binding energy when asymmetry is strong (Fig. \ref{Fig 6: MDCs at node}(b)).  Using a linear background for each MDC better reproduces the peak position (Fig. \ref{Fig 6: MDCs at node}(c)).

\begin{figure*}[!]
\includegraphics [type=jpg,ext=.jpg,read=.jpg,clip, width=4.5 in]{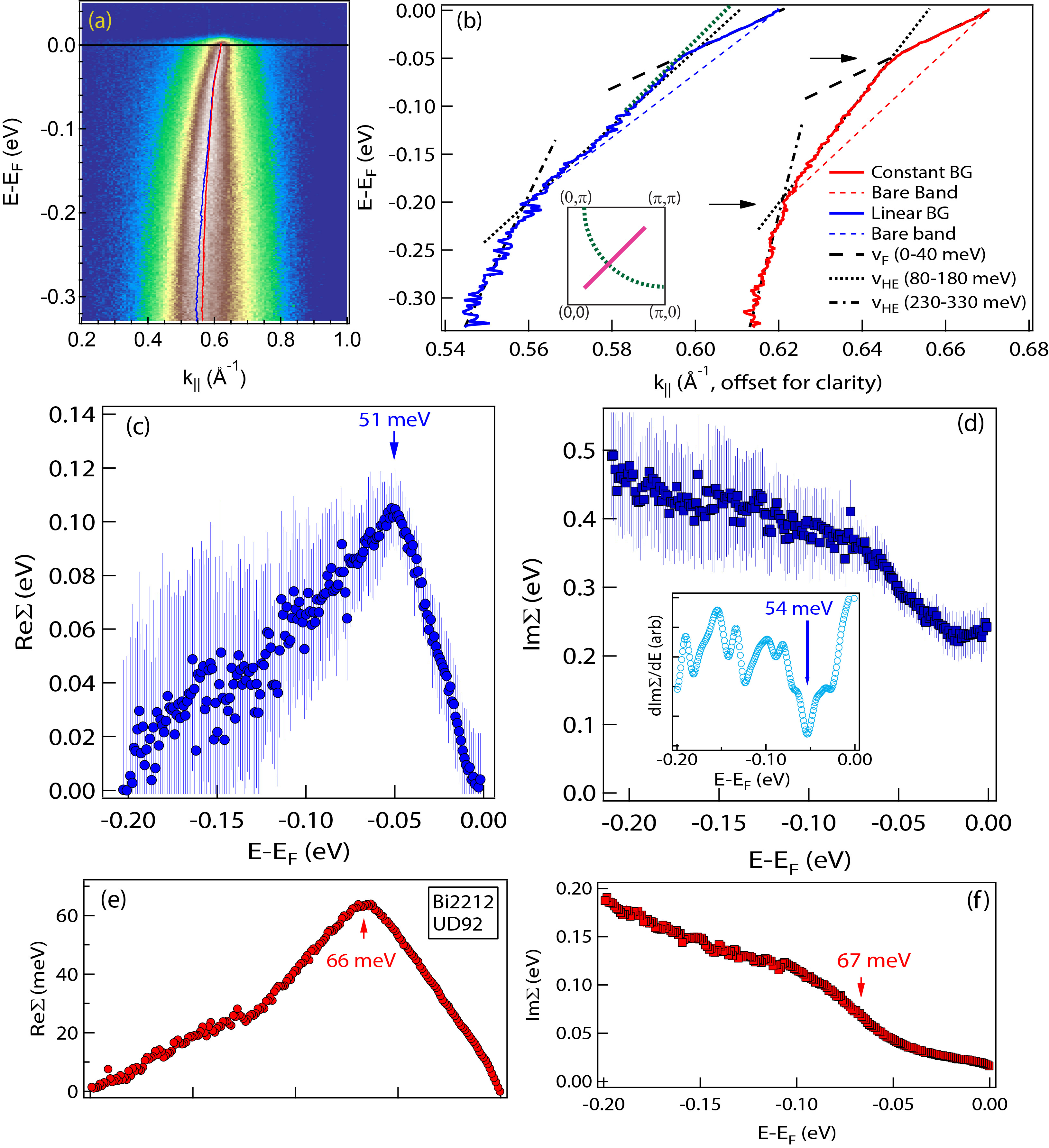}
\centering
\caption[MDC analysis comparison and self energies]{\label{Fig 7: MDC analysis and self energies} (a)-(b) Nodal MDC analysis using constant (red) and linear (blue) background.  (a) Image plot with MDC peak position from both fitting schemes.  Background EDC has been subtracted from entire image. (b) MDC peak position dispersions, offset horizontally for clarity.  Arrows mark key dispersion anomalies.  Inset shows geometry for nodal cuts. (c) Real part of self energy, Re$\Sigma$, derived using bare band indicated in (b), as discussed in text.  Linear-background fitting  (Fig. \ref{Fig 6: MDCs at node}(c)) was used. Arrow marks peak position.  (b) Imaginary part of self energy, Im$\Sigma$. Inset: derivative Im$\Sigma$ (smoothed) with respect to energy.  Arrow marks extremum. (e)-(f) Re$\Sigma$ and Im$\Sigma$ for Bi-2212 with \textit{T}$_c$=92K and p$=$0.14.  Arrows mark peak of Re$\Sigma$ and inflection of FWHM.  Bi-2212 data were taken at 10K and 3 meV energly resolution using 7eV photons and a SES2002 analyzer.}
\end{figure*}

Fig. \ref{Fig 7: MDC analysis and self energies}(b) shows MDC-derived dispersions at the node using both a constant (red) and constant$+$linear (blue) background.  Both methods indicate dispersion anomalies near 50 meV and 200 meV, and they yield comparable low-energy dispersions ($<$50 meV). The former kink is ubiquitous in all cuprates, but the latter is slightly unusual because the high-energy anomaly is generally observed at higher binding energy in hole-doped cuprates \cite{HierarchyManyBodyInteractions:Meevasana}.  The lower energy kink will be the focus of the remainder of the discussion.  Using a constant background yields steeper dispersions at higher binding energy, which is not physically correct because this fit does not capture the maximum of each MDC.  The Fermi velocity (v$_F$) is found to be similar to other cuprates\cite{UniversalNodalvF}, disregarding a potential contribution from the very low energy kink ($\omega$$<$10meV) which is not accessible in this experiment \cite{Vishik:LEKink}.  The velocity at higher energy (v$_{HE}$, 80-180 meV) is strongly dependent on the fitting scheme, with a constant background giving a slope that is 25$\%$ larger.  This discrepancy can lead to an overestimation of the mass renormalization at the $\approx$50meV kink.  A linear background yields v$_F$ (0-40 meV)$=2.008 \pm 0.002$ eV$\AA$ and v$_{HE}$ (80-180 meV)$= 3.956 \pm 0.064$ eV$\AA$ giving a mass enhancement factor $1+\lambda$$\approx$v$_F$$/$v$_{HE}$$=$1.97 $\pm$ 0.09.  %This is comparable to other cuprates near optimal doping \cite{UniversalNodalvF}.

%In photoemission, the interaction between a photon and an electron of momentum \textbf{p} is given by $\hat{H}_{int}=-\frac{e}{2mc}(\textbf{A}\cdot\textbf{p}+\textbf{p}\cdot\textbf{A})=-\frac{e}{mc}\textbf{A}\cdot\textbf{p}+\frac{\textit{i}e\hbar}{2mc}\nabla\cdot\textbf{A}$ where the latter expression makes use of the commutation relation $[\textbf{p},\textbf{A}]=-\textit{i}\hbar\nabla\cdot\textbf{A}$.  Usually, \textbf{A} is assumed to be constant, and thus $\nabla\cdot\textbf{A}=0$, but if this assumption does not hold, asymmetric lineshapes can result \cite{Hansen:SurfaceFields}.

Many-body interactions in materials are captured in the self energy, $\Sigma(\omega)$=Re$\Sigma(\omega)$+\textit{i}Im$\Sigma(\omega)$, where Re$\Sigma(\omega)$ and Im$\Sigma(\omega)$ describe interaction-induced corrections to the band dispersion and quasiparticle lifetime, respectively.  Both quantities are available from ARPES data \cite{ARPES_Review,Kordyuk:SelfEnergyAnalysis}, and because Re$\Sigma(\omega)$ and Im$\Sigma(\omega)$ are Kramers-Kronig related, many-body effects show features in both quantities.  Most analyses, including this one, assume that $\Sigma$ is independent of k normal to the FS.  Fig. \ref{Fig 7: MDC analysis and self energies}(c) shows the real part of the self energy, Re$\Sigma$,  at the node, approximated by subtracting an assumed linear bare band (blue dashed line in Fig. \ref{Fig 7: MDC analysis and self energies}(b)) from the measured MDC dispersion (blue solid line).  This quantity is peaked at 51 meV.  The imaginary part of the self energy, Im$\Sigma$, is approximated by the MDC FWHM multiplied by the slope of the assumed bare band.  Im$\Sigma$ shows an inflection point at a similar energy, affirming data quality and confirming a genuine many-body effect, electron-boson coupling, at that energy. There is a small upturn in Im$\Sigma$ near E$_F$, but this is within the error bars, and likely not significant. There was some sample-to-sample variation in the kink energy, with another good cleave showing a kink energy as high as 58 meV.  Interestingly, analysis of optical conductivity data in Hg1201 yielded a bosonic "glue" energy between 50 and 60 meV, consistent with the energy of this kink \cite{vanHeumen:OpticalDeterminationEBosonCouplingHg1201}.  In Fig. \ref{Fig 7: MDC analysis and self energies}(e)-(f), comparisons are made to Bi-2212 with a similar \textit{T}$_c$ (UD92, p$=0.14$).  The magnitude of Re$\Sigma$ is smaller in Bi-2212, suggesting stronger electron-boson coupling in Hg1201.  Additionally, the kink energy is larger in Bi-2212, with Re$\Sigma$ peaking at 66 meV.  Fig. \ref{Fig 5: Momentum dependence of gap} indicates that the superconducting gap in Hg1201 is comparable to that in Bi-2212, which suggests that bosonic modes of different energy are responsible for the kinks in the two compounds.  For further comparison, optimally doped Bi2201 (\textit{T}$_c$=33K, $\Delta$=15 meV) has multiple distinct contributions to the nodal kink, and the most prominent features appear at 70 meV and 41 meV \cite{Kondo:LEKink_Bi2201}.

%Energy range of renormalization based on all samples
The origin of the 50-80 meV nodal kink in cuprates is still a topic of debate with some explanations favoring a phononic origin \cite{MultipleModes:XJ, Devereaux:AnisotropicEph} and others favoring a magnetic origin \cite{Kaminski:kinkResonanceMode,Borisenko:ResonanceMode}.  We will consider the former first.  For optical phonons, the kink will appear at an energy $\Omega$$+$$\Delta$, where $\Omega$ is the phonon energy, because the electron-phonon coupling vertex is generally non-zero for momentum transfer \textbf{q}$=$\textbf{k$_{antinode}$}-\textbf{k$_{node}$} \cite{ElectronPhonon:Sandvik,Devereaux:AnisotropicEph}.  Thus, given a 39 meV extrapolated antinodal superconducting gap, the kink observed in Hg1201 between 50-58 meV implies an optical phonon between 11-19 meV.  Ref. \onlinecite{dAstuto:PhononHg1201} shows in-plane optical phonon branches dispersing between 8 meV and 14 meV together with an enhanced calculated phonon density of states at similar energy.  Additionally, Raman spectroscopy reports Raman-active phonons at 9 meV (Ba, E$_g$ symmetry) and near 20 meV (Apical oxygen, E$_g$; Ba, A$_{1g}$) \cite{Krantz:RamanActivePhononsHg1201,Zhou:raman1996}.  These are potential candidates for the nodal kink in Hg1201, though we note that strong coupling to low-energy optical phonons is not supported by ARPES data on other cuprates.  In Bi-2212, the nodal kink near 70 meV, is often attributed to an oxygen B$_{1g}$ mode at $\approx35$ meV. The energy of this mode is inconsistent with the difference between the kink energy and gap energy in Hg1201, and additionally, single-layer cuprates are not expected to show electron-phonon coupling to first order for this mode \cite{JohnstonVernay:SystematicStudye-ph}.  In Hg1201, the lack of a neutral cleavage plane and resultant surface electric fields presents an added complication.  Surface electric fields could cause electrons near the surface to couple to phonons which are not Raman-active from symmetry arguments, and it could also affect the strength of electron-phonon coupling \cite{JohnstonVernay:SystematicStudye-ph}.  Future information about the precise surface termination from scanning tunneling microscopy could clarify how to interpret these data in terms of electron-phonon coupling.
%This mode involves c-axis displacements of planar oxygen.  In single-layer cuprates, electrons should not couple to this mode to first order because the CuO$_2$ plane is a mirror plane \cite{JohnstonVernay:SystematicStudye-ph}.  The oxygen B$_{1g}$ mode involves c-axis displacements of planar oxygen, coupling to it, as seen by a surface spectroscopy, can be affected by a polar cleaved surface.  This may alternately explain the differing kink energies in Hg1201 and Bi-2212.

Second, we explore the possibility of a magnetic origin of the kink in Hg1201.  In neutron scattering, a magnetic collective mode has been observed below the pseudogap temperature (T$<$ T*), which disperses from 52 meV to 56 meV between \textbf{q}=(0,0) and \textbf{q}=(0.5,0.5) \cite{Li:MagneticModeHg1201,Li:TwoIsingLikeMagneticExcitations}.  This energy is intriguingly similar to the energy position of the nodal kink observed with ARPES. If the two phenomena are indeed related, it would imply that nodal electrons couple to this magnetic collective mode without effects due to the superconducting gap elsewhere in momentum space.  This can happen if coupling is strongly peaked at \textbf{q}$=$0 (forward scattering) or at a wavevector connecting two nodal points.  Further doping, temperature, and momentum-dependent investigation is needed to clarify the origin of the nodal kink seen in Hg1201. %Further investigation of the kink away from the node can clarify the relationship between the collective mode seen by neutron scattering and ARPES spectra in Hg1201, though preliminary analysis of dispersions in Fig. \ref{Fig 1: Crystal structure and hn dep}(a)-(e) indicates that the kink energy does not disperse away from the node, consistent with near-optimally-doped Bi-2212\cite{Cuk:CouplingB1g}.  %Studying the momentum dependence of the kink in Hg1201 indicates that the kink energy does not disperse away from the node (Fig. \ref{Momentum dependence of kink}).  The kink energy appears similar for all momenta, suggesting that at every momentum, the maximum of the \textit{d}-wave gap--not its local value--is the relevant quantity.

%\begin{figure}[h!]
%\includegraphics [type=eps,ext=.eps,read=.eps,clip, width=5 in]{./Hg1201/momentum_dependence_kink}
%\centering
%\caption[Momentum dependence of kink]{\label{Momentum dependence of kink} MDC analysis for data in Fig. \ref{Fig 3: Near-nodal cuts Hg1201}.  Dispersions offset horizontally for clarity.  Red horizontal dashed line is approximate kink energy, as quantified by the energy where dispersion deviates from low energy linear dispersion.  Purple and Black dashed lines are offset from red by 14 meV and 20 meV, respectively, reflecting the measured gaps at 33$^\circ$ and 27$^\circ$, respectively. }
%\end{figure}

%prominent bosons in Hg1201 and which ones may be associated with kink
%comparison to other cuprates (kink energy, renormalization energy, gap)
%future studies
%Maybe analyze T-dep data

\section{Conclusions}
Despite its lack of a natural cleavage plane, Hg1201 is a crucial compound to explore with ARPES because of its structural simplicity and the wealth of high-quality data obtained with other probes.  The present work describes our technical experimental progress and measurement of basic electronic properties available in ARPES: superconducting gap, nodal kink energy, and band renormalization.

Future studies may be aided by a more controlled method to cleave the samples \cite{Takayuki:CleavTinySamples,Mansson:OnBoardSampleCleaver}, by better control of the surface termination, and by finding experimental conditions where the antinode has adequate cross section.  Experiments with higher photon energy (h$\nu$$>$100 eV) may be promising to achieve the latter goal \cite{Noh:PdCoO2ARPES}.

Ultimately, studies of Hg1201 may be very fruitful in addressing the question of 'what causes a high \textit{T}$_c$?' via comparisons to other single layer cuprates.  A great body of ARPES studies exists for LSCO and Bi-2201, the 'low \textit{T}$_c$' single-layer cuprates \cite{ARPES_Review,Hashimoto:LSCO_doping_dep,yoshida:LSCOGapFunction2009,Hashimoto:DopingEvolutionBi2201,Kondo:CompetitionSCPG_Nature_2009}.  In contrast, there have been fewer ARPES studies on Tl$_2$Ba$_2$CuO$_{6+\delta}$ (Tl2201) \cite{Plate:OD_Tl2201_ARPES,Lee:Tl_cuprate,Palczewski:singleLayerCupratesARPES_Tl2201} and none on Hg1201 prior to the present work.  Understanding the electronic structure of these 'high \textit{T}$_c$' single layer cuprates is crucial for discerning the ingredients for higher \textit{T}$_c$ .   With comprehensive studies of Hg1201 and Tl2201, new and non-trivial features may be found in the electronic structure which distinguish the low \textit{T}$_c$ compounds from the high \textit{T}$_c$ ones.

\begin{acknowledgments}
The authors acknowledge helpful discussions with Sri Raghu, Edbert Jarvis Sie, and Fahad Mahmood. SSRL is operated by the DOE Office of Basic Energy Science, Division of Chemical Science and Material Science. This work is supported by DOE Office of Science, Division of Materials Science, with contracts DE-FG03-01ER45929-A001 and DE-AC02-76SF00515.  The work at the University of Minnesota was supported by the Department of Energy, Office of Basic Energy Sciences under Award No. DE-SC0006858. N.B. acknowledges support though a Marie Curie Fellowship and European Research Council (Advanced Grant Quantum Puzzle, no. 227378).
\end{acknowledgments}

\bibliography{references}
\end{document}